\documentstyle[epsfig,longtable]{aipproc}

\begin{document}
\title{Rotational Instabilities and Centrifugal Hangup}

\author{Kimberly C. B. New$^*$ and Joan M. Centrella$^{\dagger}$}
\address{$^*$Los Alamos National Laboratory, Los Alamos, New Mexico 87545\thanks
{A portion of this work was performed under the auspices of the U.S. Department of Energy by Los Alamos National Laboratory under contract W-7504-ENG-36.}\\
$^{\dagger}$Drexel University, Philadelphia, Pennsylvania, 19104}

\maketitle

\begin{abstract}
One interesting class of gravitational radiation sources includes rapidly
rotating astrophysical objects that encounter dynamical instabilities.
We have carried out a set of simulations of rotationally induced instabilities
in differentially rotating polytropes.  An $n$=1.5 polytrope with the Maclaurin
rotation law will encounter the $m$=2 bar instability at $T/|W| \gtrsim 0.27$.
Our results indicate that the remnant of this instability is a persistent 
bar-like
structure that emits a long-lived gravitational radiation signal.  Furthermore,
dynamical instability is shown to occur in $n$=3.33 polytropes with the
$j$-constant rotation law at $T/|W| \gtrsim 0.14$.  In this case,
the dominant mode
of instability is $m$=1.  Such instability may allow a centrifugally-hung core
to begin collapsing to neutron star densities on a dynamical timescale.  If it
occurs in a supermassive star, it may produce gravitational radiation
detectable by LISA.
\end{abstract}

\section*{Introduction}
%
%
%
%

One interesting class of gravitational radiation sources includes rapidly
rotating astrophysical objects that encounter dynamical instabilities.
Linear stability analysis has shown that rapidly rotating bodies may experience
global deformations due to the growth of unstable azimuthal modes
$e^{\pm im\phi}$ \cite{chan69,tass78}.
The mode numbers $m$ describe the shape of the induced
deformation.  For example, an $m$=1 mode could result in the development of a
one-armed spiral or a simple translation; an $m$=2 mode may produce a
bar-shaped distortion; an $m$=3 mode, a triangular distortion; and so on.
The onset of such an instability occurs when an object's ratio of rotational
kinetic energy $T$ to gravitational potential energy $W$, $\beta=T/|W|$,
exceeds a critical value $\beta_{crit}$.

Both secular and dynamical varieties of these instabilities may exist.
A dynamical instability is driven by gravitational and hydrodynamical
forces and develops on the order of the rotation period of the system.
A secular instability is driven by a dissipative mechanism, such as
viscosity or gravitational radiation, and develops on the timescale of
that mechanism (which can be many rotation periods).  In this paper, we
focus on the numerical study of dynamical instabilities, since their
development can be followed in a reasonable amount of computational time
with explicit hydrodynamical simulations.

There are several types of astrophysical objects that may encounter these
rotational instabilities.  A star that accretes matter and angular momentum
from a binary companion may be spun up to rapid rotation \cite{schutz89}.
A second example is a centrifugally hung stellar core or ``fizzler''
\cite{thorne96,tohl84,shli76}.  The formation of a fizzler begins when the
core of a massive star depletes its nuclear fuel and begins to collapse to
neutron star densities.  If the core was rotating initially,
conservation of angular momentum requires that the core spin up as it
collapses.  This spin up could actually halt the collapse if the centrifugal
force increases to the point where it overcomes the inward gravitational push.
The results would be a rapidly rotating, partially collapsed stellar core.
Another example is a cooling supermassive star (mass $>10^6 M_{\sun}$)
that also spins up as it contracts \cite{nesh00,bash99}.  Finally, if
the merger of a compact binary does not result in an immediate collapse
to a black hole, the remnant will be a rapidly rotating compact object
\cite{lash95}.  The nonaxisymmetric deformations induced by rotational
instabilities could result in relatively strong gravitational radiation
emission from these rapidly rotating objects.

We have performed Newtonian hydrodynamics simulations of dynamical
rotational instabilities in two different types of objects
\cite{cnlb00,nct00}.  The
much studied $m$=2 bar mode is the strongest of the set of global
instabilities encountered by an $n$=1.5 polytrope (for which the equation
of state gives the pressure $P$ in terms of the density $\rho$ as
$P=K\rho^{(1+1/n)}$, where $K$ is the polytropic constant) with the Maclaurin
differential rotation law.  This instability sets in when $\beta \gtrsim
0.27$.  Our results indicate that dynamical instabilities also occur
in objects with much lower values of $\beta$.  In fact, $n$=3.33 polytropes
with the $j$-constant rotation law encounter a dominant $m$=1 instability
when $\beta$ reaches $0.14$.  Simulations of the bar mode instability and
the $m$=1 instability will be discussed in the following sections.

\section*{The Bar Mode Instability}

There has been a discrepancy in the outcome of previous hydrodynamical
simulations of the dynamical bar mode.  Namely, simulations carried
out by Centrella's group showed that an object deformed by the bar mode
would become axisymmetric again after a short interval \cite{shc96,hoce94}.
This is in contrast to the simulations performed by several other groups,
which resulted in bar-shaped final configurations
(e.g., \cite{new96,pick96,duri86}).

The degree of asymmetry in the final configuration is important because
an axisymmetric object (rotating about its short symmetry axis) will not
emit gravitational radiation.  If the simulations performed by Centrella's
groups's simulations are correct, the instability would produce only a short
burst of radiation.
However, if the object retains a bar-like structure, as the other simulations
seem to indicate, it will go on emitting gravitational radiation, producing
a longer-lived signal that would be easier to detect.

This discrepancy is illustrated by the gravitational waveforms shown in
Figures 1 and 2.  The waveform in Figure 1 is from a simulation performed by
Centrella's group \cite{shc96},
with a finite-difference hydrodynamics (FDH) code
developed at Drexel University (hereafter called the ${\cal D}$ code).
This waveform is burstlike; the amplitude dies off as the object
loses its bar-like shape.  The waveform in Figure 2 is from a simulation
performed by New \cite{new96}, with a FDH code developed at Louisiana
State University (hereafter called the ${\cal L}$ code).  This waveform
rings for the duration of the simulation because the object retains its
bar-shaped structure.

\begin{figure}
\centerline{\epsfig{file=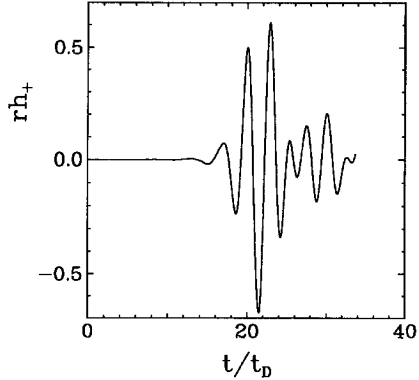,height=2in}}
\caption{The $h_{+}$ polarization of the gravitational waveform from
a ${\cal D}$ code bar mode simulation, as viewed by an
observer located along the rotation axis at a distance $r$ from the system.
The value $rh_+$
has been normalized to $R^{-1}c^{-4}M^2G^2$; the time is normalized
to the dynamical time $t_{D}=[R^3/(GM)]^{1/2}$.}
\end{figure}
\begin{figure}
\centerline{\epsfig{file=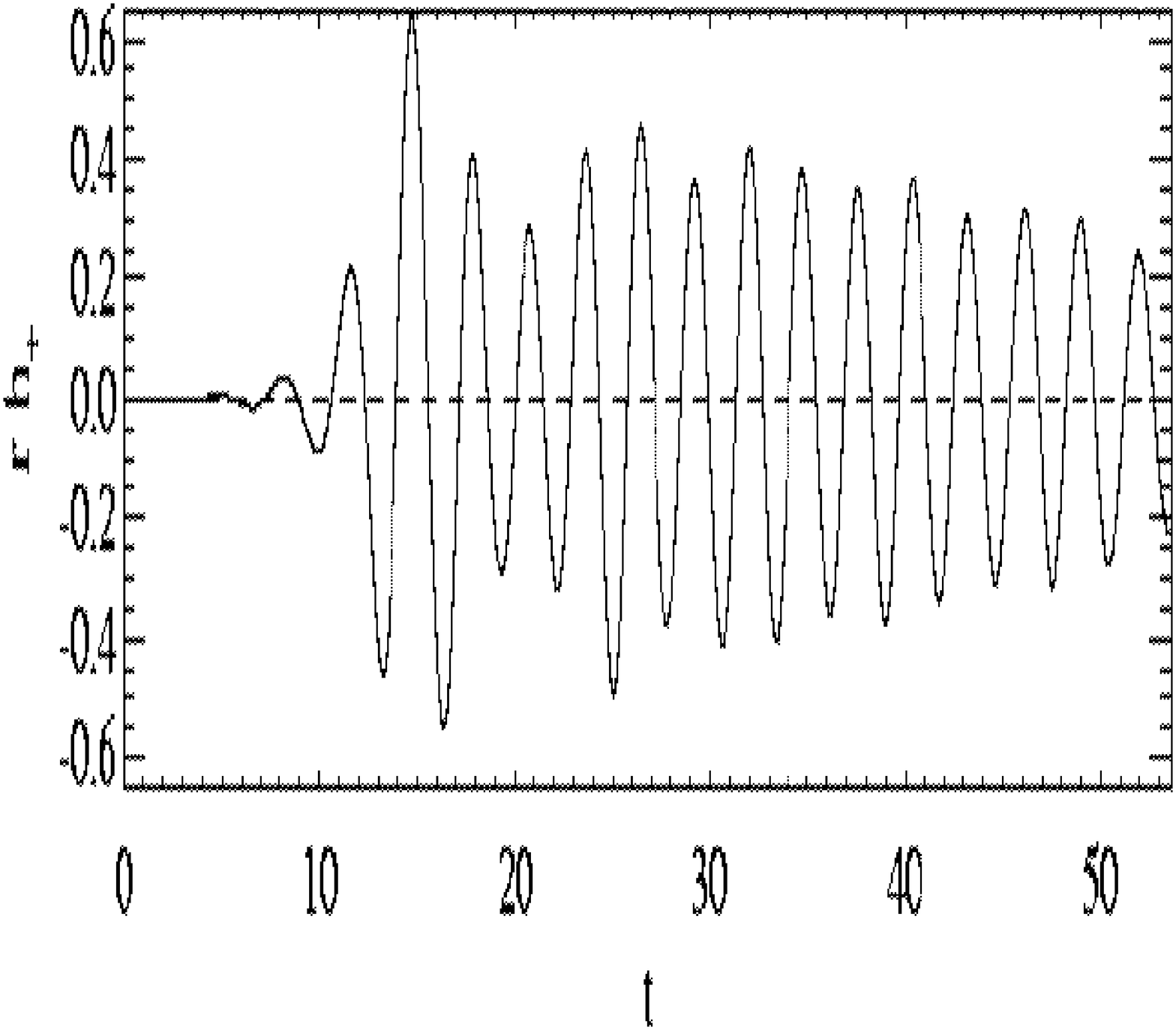,height=2in}}
\caption{Same as Figure 1, for a simulation performed
with the ${\cal L}$ code.}
\end{figure}

Again, the resolution of this discrepancy is important because if the
gravitational radiation signal persists, the total accumulated amplitude
could make it easier to detect.  The characteristics of New's waveform
are such that if the star's initial mass and radius are $M=1.4 M_{\sun}$
and $R=36\,{\rm km}$, and it is located at a distance of 20 Mpc, the maximum
amplitude is $h_{max} \sim 10^{-23}$ and the frequency $f_{gw}$ ranges from
$\sim 600-670\,{\rm Hz}$.  The scaling is such that $h_{max} \propto R^{-1}$
and $f_{gw} \propto R^{-3/2}$.  This signal could possibly be detected with
advanced interferometers like LIGO II.

In a recent paper \cite{nct00}, we demonstrate that a good deal of
progress has been made in resolving the discrepancy among the outcomes
of the previous simulations of the bar instability.  In an attempt to
get at the root of the problem, we examined the differences between
the simulations performed with the ${\cal D}$ and ${\cal L}$ codes.
One of the main differences is that New's simulation with the ${\cal L}$
code \cite{new96} used a condition called $\pi$-symmetry.  The $\pi$-symmetry
condition enforces periodic azimuthal symmetry and thus allows the
azimuthal resolution to be doubled as the solution only needs to be
evolved from $0$ to $\pi$.  This symmetry condition does not allow the
growth of odd modes in a simulation.  However, the use of $\pi$-symmetry
in a bar mode simulation seems justified by analytic perturbation anaylsis,
which indicates that odd modes will not grow if $\beta < 0.32$ \cite{toma98}
($\beta$=0.30 in the initial models used by \cite{shc96,hoce94,new96}). 

In order to investigate if the use of $\pi$-symmetry was responsible for
differences in the outcomes of the ${\cal D}$ and ${\cal L}$ codes'
bar mode simulations, we reran the ${\cal L}$ code simulation with
the $\pi$-symmetry condition turned off (hereafter referred to as
simulation ${\cal L}$1).  The ${\cal L}$1 simulation started
with the same axisymmetric
initial model used in the previous simulations \cite{shc96,new96}.
It is constructed in hydrostatic equilibrium with Hachisu's Self-Consistent
Field (SCF) method \cite{hach86}.  It has an $n$=1.5 polytropic equation of
state and is differentially rotating, with an angular momentum
distribution identical to that of a Maclaurin spheroid \cite{boos73}.
It is a highly flattened, rapidly rotating object with $\beta \sim 0.3$
(recall $\beta_{crit} \approx 0.27$).

Equatorial density contours from the last half of the ${\cal L}$1 simulation
are shown in Figure 3.  As the final frame indicates, the object appears
rather symmetric at the end of the run.  The resulting waveform
(Figure 4) reflects the object's return to symmetry.  This waveform
looks much more like the burstlike waveforms of Centrella's group
\cite{shc96,hoce94} (see, e.g., Figure 1).

\begin{figure}
\centerline{\epsfig{file=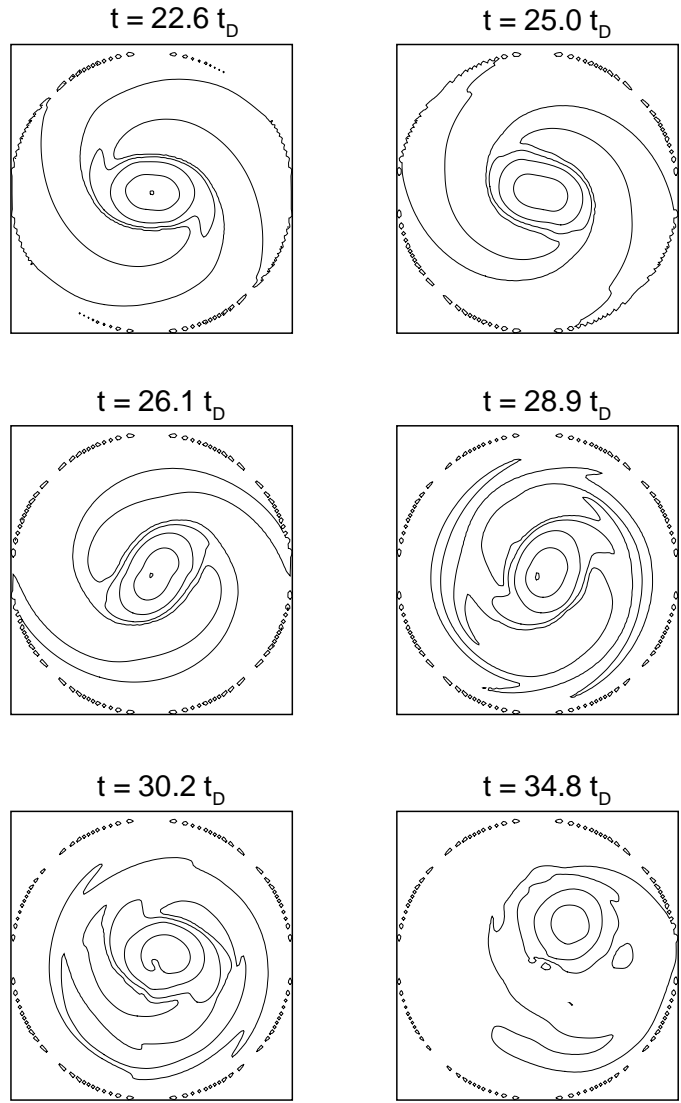,height=6.in}}
\caption{Density contours in the equatorial plane for the later
stages of the ${\cal L}$1 simulation are shown.  The maximum
density has been normalized to unity at $t$=0 and the contour
levels are at 0.5, 0.05, 0.005, and 0.0005.}
\end{figure}

\begin{figure}
\centerline{\epsfig{file=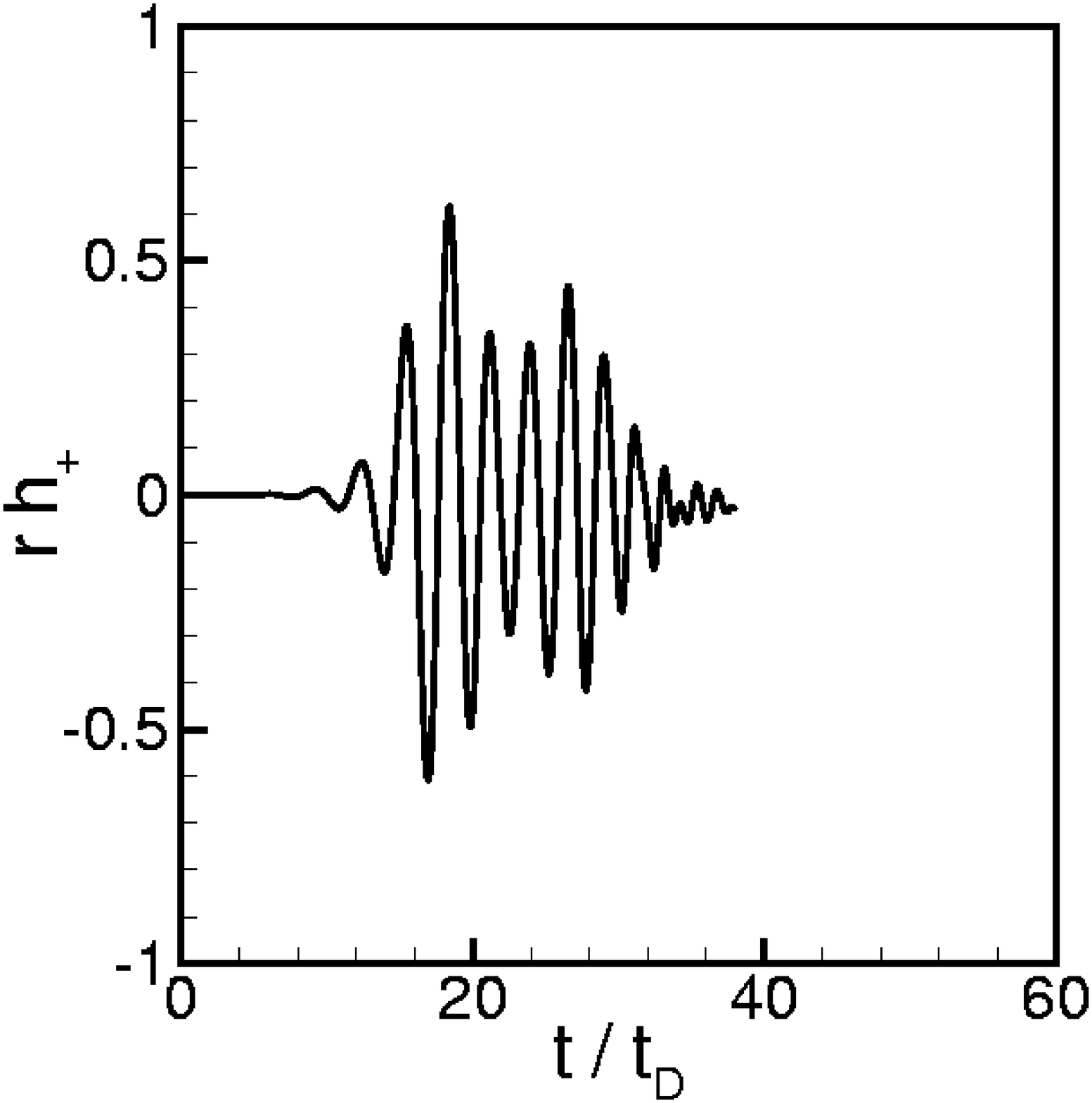,height=2.5in}}
\caption{The gravitational waveform from the ${\cal L}$1 simulation
is shown.  The normalizations are identical to those used in Figure 1.}
\end{figure}

Thus the ${\cal L}$1 run appears to be in better 
agreement with the ${\cal D}$
code simulations.  But the important question is whether the result
on which they agree is the ``physical'' result.  That is, can we with
confidence tell the gravitational wave detection community that a star
that encounters the bar instability will emit only a burst of radiation?

The answer to that question is {\it no}.  This is because of an effect that
appears late in the ${\cal L}$1 simulation.  At about the time that the
object loses its bar-like structure, the center of mass of the object moves
off the center of the grid.  This can be seen in the final frame of
Figure 3.  In the absence of external forces, the center of mass should
not move.  This is thus a nonphysical result and must be due to a shortcoming
in the numerics of the ${\cal L}$ code.

Before we could say for certain that an axisymmetric configuration is the
correct remnant of the bar instability, we needed to find a way to
minimize the center of mass motion and see how this affected the outcome
of the evolution.  Further testing indicated that increasing the radial
resolution of the simulation does indeed delay the onset of the motion
of the center of mass.

We ran the bar mode simulation with the ${\cal L}$ code once again,
this time with double the radial (and axial) resolution used in
the ${\cal L}$1 run (this high resolution run will be referred
to as ${\cal L}$2).  In the ${\cal L}$2 evolution, the bar-like
structure is maintained for about 15 more dynamical times (where
$t_{D}=(R^3/GM)^{1/2}$) than in the ${\cal L}$1 run.  The waveform
from this run, shown in Figure 6, has a correspondingly longer duration.
Doubling the resolution delayed the onset of the center of mass motion,
but it did not prevent it.  Equatorial density contour plots from the
latter portion of the ${\cal L}$2 run are shown in Figure 5.  As is
evident in the final frame, the bar-like structure (and waveform) decay
at the time that the center of mass starts to move.

\begin{figure}
\centerline{\epsfig{file=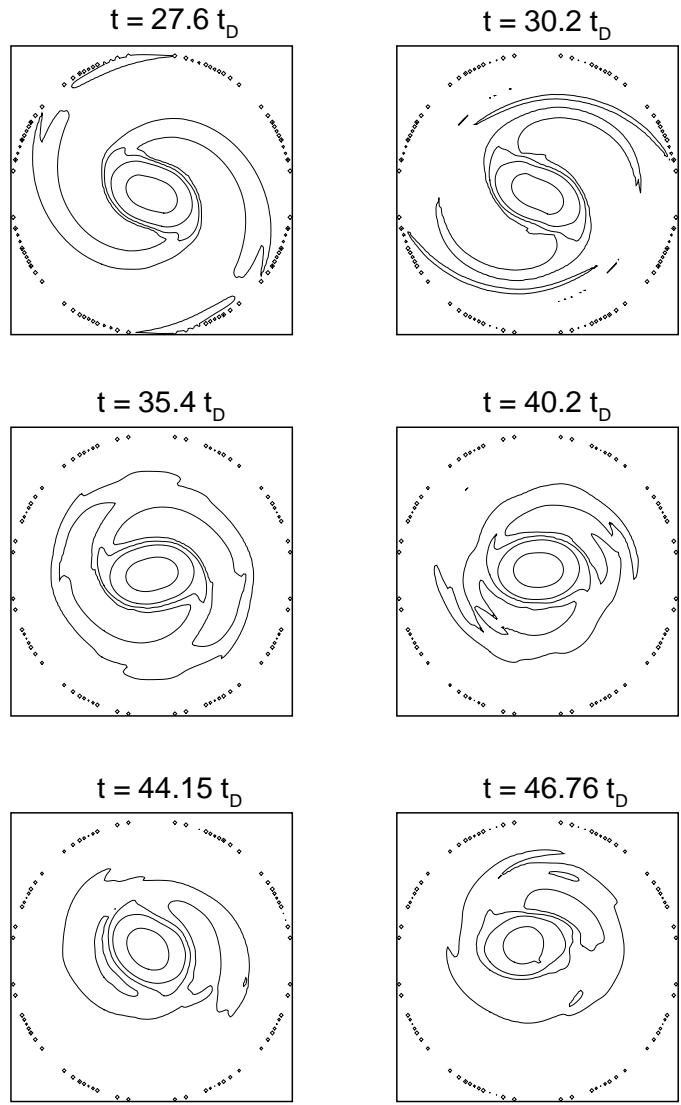,height=6.in}}
\caption{Density contours in the equatorial plane for the later
stages of the ${\cal L}$2 simulation are shown.  The contour
levels are the same as those shown in Figure 3.}
\end{figure}
\begin{figure}
\centerline{\epsfig{file=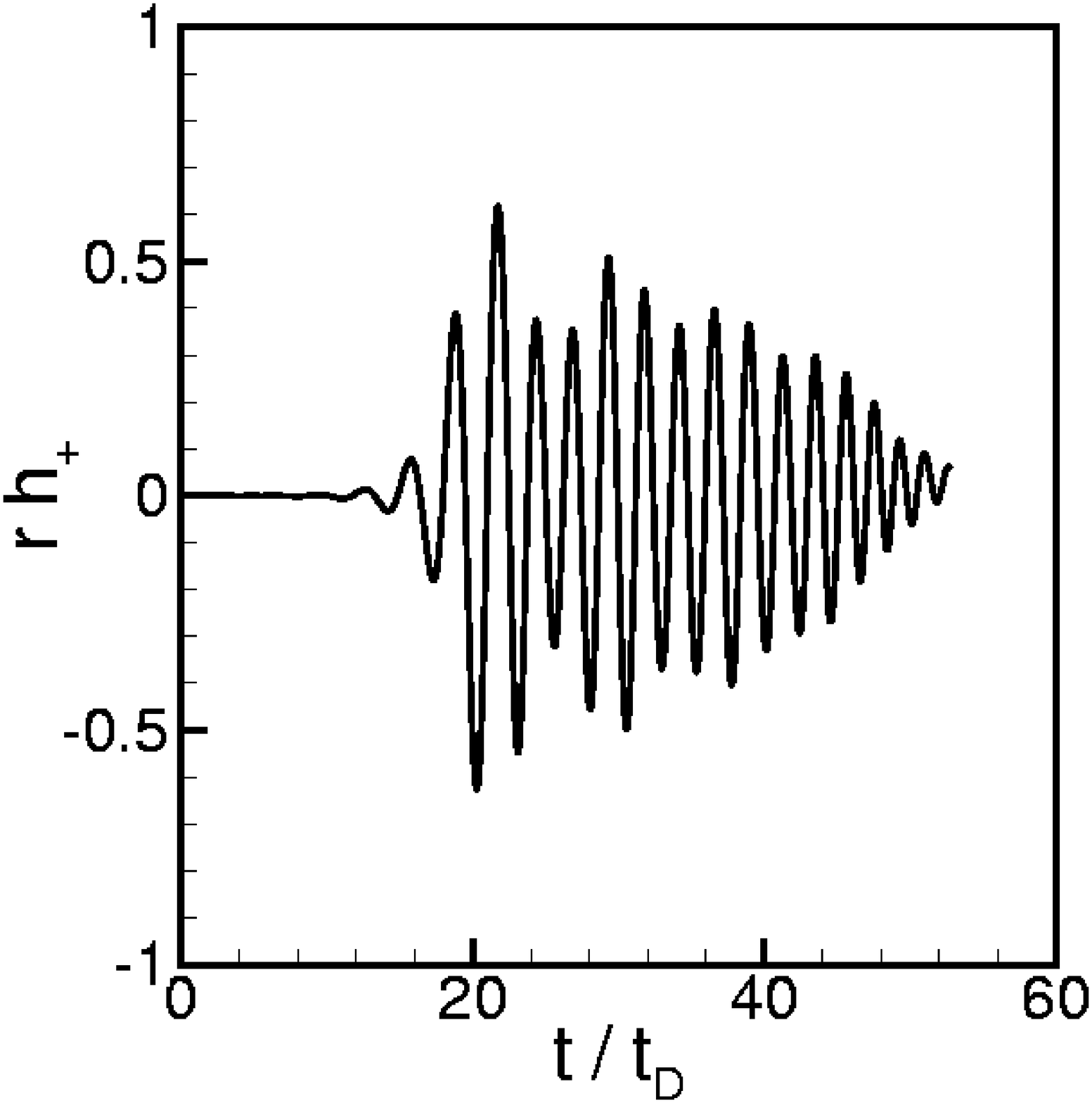,height=2.5in}}
\caption{The gravitational waveform from the ${\cal L}$2 simulation
is shown.  The normalizations are identical to those used in Figure 1.}
\end{figure}

We have demonstrated that the longer the center of mass motion is suppressed,
the longer the bar-like structure is maintained.  Since the center of mass
motion is unphysical, the decay of the bar is as well.  Thus a simulation
with very high resolution, or better finite-differencing algorithms that
did not develop center of mass motion, should produce a long-lived
nonaxisymmetric structure.  A recent paper by Brown confirms this
\cite{brown00}.  The FDH PPM hydrocode Brown used to simulate the bar
instability explicitly conserves linear momentum, thus preventing center
of mass motion from developing.  The outcome of Brown's simulation was
indeed a persistent bar.

Recall that this is the outcome of the ${\cal L}$ code's simulation
with $\pi$-symmetry \cite{new96}.  This symmetry condition prevents
the growth of any center of mass motion because it will not allow
an $m$=1 mode to grow.  Our results and the recent paper by Brown
\cite{nct00,brown00} indicate that the physically accurate outcome
of the bar instability in this object is a persistent bar-like configuration
that emits gravitational radiation over many cycles and thus may be capable
of producing a detectable signal.

\section*{Instability at Low $\beta$}

We have recently also investigated dynamical rotational instabilities in 
objects with lower values of $\beta$ \cite{cnlb00}.  This study was motivated
by the fact that centrifugally hung stellar cores, or fizzlers, are
likely to have $\beta < 0.27$ and thus will not encounter the particular
bar mode instability that was discussed in the preceeding section
\cite{tohl84,zwmu97,ermu85}.

Previous authors have identified dynamical instabilities in toroidal
configurations at values of $\beta \gtrsim 0.1$ \cite{wood94,toha90}.
Thus we decided to investigate the stability properties of spheroidal
configurations with off-center density maxima (thinking
they may behave more like tori).  Our goal was to determine whether
spheroidal configurations, representative of fizzlers, could
encounter instabilities at lower values of $\beta$.

The initial models for this stability study were constructed with
Hachisu's SCF method.  An $n$=3.33 polytropic equation of state was
used and is representative of a partially collapsed stellar core.
Because we wanted to study models with off-center density maxima,
we needed to use a rotation law other than the Maclaurin law.  The
$j$-constant law does produce off-center density maxima in $n$=3.33
polytropes.  The angular velocity distribution $\Omega(\varpi)$ given
by this law is
\begin{equation}
\Omega^2=\frac{j_{0}^{2}}{(d^2+\varpi^2)^2},
\end{equation}
where $\varpi$ is the cylindrical radius and $d$ is an arbitrary constant.
As $d$ approaches zero, the specific angular momentum approaches the
constant $j_0$.  We constructed models with $d$=0.2.  Figure 7 displays
meridional density contour plots of four of the equilibrium models
we constructed.  Off-center density maxima appear as $\beta$ is increased.

\begin{figure}
\centerline{\epsfig{file=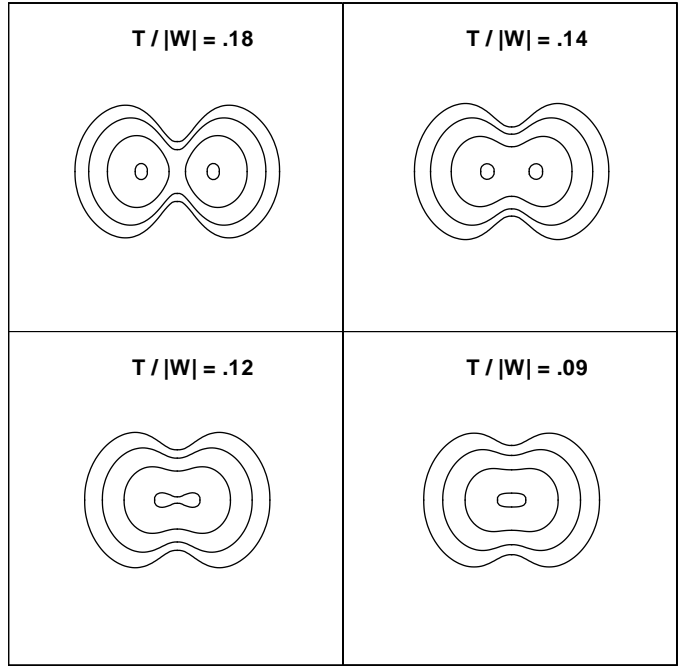,height=3.5in}}
\caption{Density contours in the meridional plane
are shown for 4 equilibrium models with $d$=0.2 and $n$=3.33.
The models are labeled with their values of $\beta=T/|W|$.  The
initial maximum
density has been normalized to unity and the contour
levels are at 0.9, 0.1, 0.01, and 0.001.}
\end{figure}

We performed simulations of these models with the ${\cal L}$ code and
with Brown's PPM hydrocode.  The ${\cal L}$ code uses a cylindrical
grid and Brown's code uses a Cartesian grid.  We have evolved all four
of the models shown in Figure 7.  The onset of instability is near
$\beta$=0.14.  Both the $\beta$=0.14 and 0.18 models are clearly
unstable (see below).  We have also run simulations of the models with
$\beta$=0.12 and $0.09$.  The $\beta$=0.12 model was run for about
40 $t_{D}$.  At the end of the simulations, the $m$=1 mode was just
starting to grow.  The $\beta$=0.09 model was run for 35 $t_{D}$
and showed no mode growth.  We plan to run longer, higher resolution
simulations to determine the value of $\beta$ at which the instability
sets in.  Note that no significant center of mass motion was seen in
these simulations.

The evolution of the model with $\beta$=0.14 exhibits a dynamical
instability that is very similar in simulations performed with both codes.
The development of the instability is shown in the equatorial density
contour plots of Figure 8 (from the ${\cal L}$ code run).  The torus
pinches off to produce a single high density region, indicative of
a $m$=1 mode.  This dense region starts to collapse at late times,
as is evident from the appearance of higher density contours.  Figures
9 and 10 show the amplitude of the modes that grow during the evolutions
performed with both codes.  The $m$=2, 3, and 4 modes grow in sequence
after the $m$=1 mode.  The pattern speeds for the $m$=1 and $m$=2 modes
are identical.  Thus the $m$=2 mode is an harmonic of the $m$=1 mode.
The constant amplitude of the $m$=4 mode in Brown's simulation is a
result of the symmetry of his Cartesian grid.

\begin{figure}
\caption{Density contours in the equatorial plane
are shown for the model with $\beta$=0.14.
The darkness of the shading increases as the density
decreases.  The contour levels are 0.01, 0.1, 0.9, 2,
and 4 times the maximum density at the initial time.}
\end{figure}

\begin{figure}
\centerline{\epsfig{file=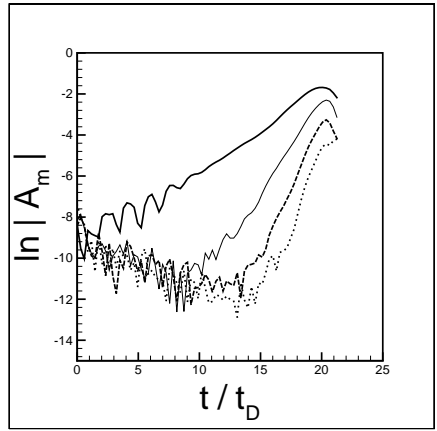,height=2.25in}}
\caption{
The growth of the amplitudes $|A_m|$ for
$m$=1 (thick solid line), $m$=2 (thin solid line),
$m$=3 (dashed line), and $m$=4 (dotted line)
is shown for the ${\cal L}$ code simulation of the $\beta$=0.14 model.
These amplitudes were
calculated in the equatorial plane for a ring with
radius $\varpi$=0.32.}
\end{figure}
\begin{figure}
\centerline{\epsfig{file=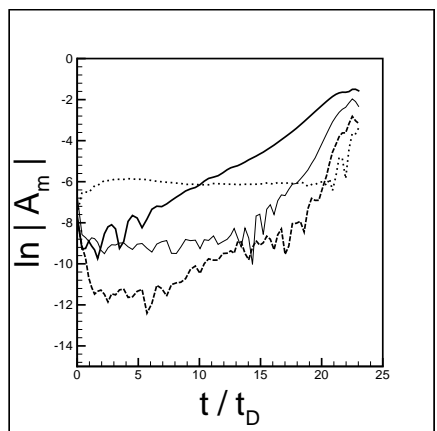,height=2.25in}}
\caption{
Same as Figure 9, for Brown's simulation.}
\end{figure}

The evolutions of the $\beta$=0.18 model also exhibit dynamical
instability.  The mode amplitude plots from these runs are shown
in Figures 11 and 12.  The instability grows more quickly in this model
than in the $\beta$=0.14 model, as expected.  The $m$=1 mode grows
at about the same rate in the simulations of both codes.  The
${\cal L}$ code run shows the $m$=2, 3, and 4 modes growing in
sequence, just like in the $\beta$=0.14 simulation; the $m$=2 mode
is again an harmonic of the $m$=1 mode.  Brown's run, however,
shows the $m$=1 and $m$=2 modes growing at the same rate.  These
modes are not harmonics in his simulation.  We plan to investigate
the differences in these results with higher resolution simulations.

\begin{figure}
\centerline{\epsfig{file=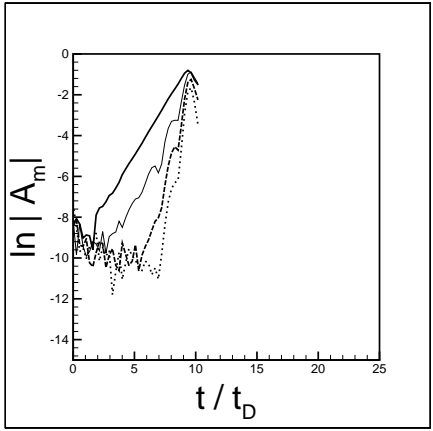,height=2.25in}}
\caption{Same as Figure 9, except that the model
has $\beta$=0.18.}
\end{figure}
\begin{figure}
\centerline{\epsfig{file=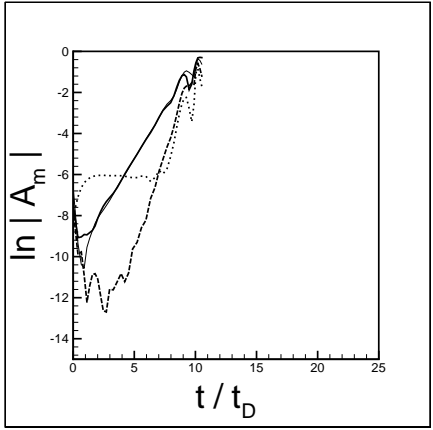,height=2.25in}}
\caption{Same as Figure 10, except that the model
has $\beta$=0.18.}
\end{figure}

Our study demonstrates that dynamical instabilities can occur in
differentially rotating polytropes with relatively low values of
$\beta$.  Note that Pickett, Durisen, and Davis
\cite{pick96} also found a dominant
$m$=1 instability in a centrally condensed configuration.  The instability
in their model set in at $\beta$=0.2. The model had an $n$=1.5
polytropic equation of state and was constructed with the so-called
$n'$=2 rotation law.  This rotation law places more angular momentum
in the outer regions of the model than does the $j$-constant law.  Their
model did not have off-center density maxima.

We plan to carry out more detailed studies to investigate the character
of these unstable modes and their properties for various values of
the polytropic index $n$ and the rotation law parameter $d$. 

If this instability occurs in centrifugally hung stellar cores, collapse
to neutron star densities may result.  Our simulations do show that the
density is increasing at the end of the unstable runs.  Further studies
are needed to determine how dense the remnant becomes and to see if the
$m$=1 mode will result in the star moving with velocities comparable to
those of actual neutron stars.

Longer runs on larger grids are needed to obtain the full gravitational
radiation waveforms.  However, we can estimate the properties of the
emission during the initial stages of the development of the instability.
For a fizzler that starts out with $M \sim 1.4 M_{\odot}$ and $R \sim 200$km,
the peak emission will occur at $f_{gw} \sim 200$Hz.  The peak amplitude
$h_{max}$ will be $\sim 10^{-24} r_{20}^{-1}$ for $\beta = 0.14$, and
$\sim 10^{-23} r_{20}^{-1}$ for $\beta = 0.18$.  Here, $r_{20}$ is
the distance to the source in units of 20 Mpc.  Emission from such unstable
cores may be detectable with advanced ground-based interferometers like
LIGO II.

An even more optimistic scenario for the detection of gravitational radiation
occurs if this type of instability is encountered by a cooling and contracting
supermassive star \cite{nesh00,bash99}.  If the star's ratio $GM/Rc^2$
is $\sim 15$ when the instability develops (a value that is approximately
appropriate for an uniformly rotating star), $f_{gw}$ will be $\sim 10^{-3}$
Hz and $h_{max}$ will be $\sim 10^{-18}r_{20}^{-1}$ for $\beta =
0.14$ and $\sim 10^{-17}r_{20}^{-1}$ for $\beta = 0.18$.
Such signals would be easily
detectable by the space-based LISA detector.

\end{document}